\documentclass[1,authoryear,12pt]{elsarticle}
\usepackage{eurosym}
\usepackage{amsmath}
\usepackage{amsfonts}
\usepackage{graphicx}
\usepackage{subfigure}
\usepackage{url}
\usepackage{xcolor}
\usepackage{calligra}
\usepackage[T1]{fontenc}
\input{tcilatex}
\journal{Mechanics of Materials, accepted for publication}

\begin{document}

\begin{frontmatter}

\title{Simulation of dislocation slip and twin propagation in Mg through coupling crystal plasticity and phase field models}
\author[L1]{Meijuan Zhang}
\author[L1]{Anxin Ma \corref{C1}}
\ead{anxin.ma@imdea.org}
\author[L1, L2]{Javier Llorca}
\address[L1]{IMDEA Materials Institute, c/ Eric Kandel 2, 28906 Getafe, Madrid, Spain}
\address[L2]{Department of Materials Science, Polytechnic University of Madrid, 28040 Madrid, Spain}
\cortext[C1]{Corresponding author. tel +34 91 549 34 22}

\begin{abstract}
A numerical strategy to simulate plastic deformation in Mg alloys including dislocation slip and twin propagation is presented. Dislocation slip is included through a crystal plasticity model which is solved using the finite element method while twin propagation is taken into account by means of a phase field model which is solved using a fast Fourier transform algorithm. The coupled  crystal plasticity and phase field equations were solved using different discretizations of the simulation domain using the same time step for both of them. The numerical strategy was used to simulate the deformation in compression of a Mg micro-pillar along the $[10\bar{1}0]$ direction. The stress-strain curve predicted by the model as well as the dominant deformation mechanisms were in agreement with the experimental data in the literature and demonstrate the viability of the strategy to take explicitly into account twin propagation during the simulation of plastic deformation of Mg alloys. Finally, an example of slip/twin interaction in polycrystals was simulated to show the capabilities of the model.  
\end{abstract}


\end{frontmatter}

\section{Introduction}
Mg alloys are currently used to manufacture non-structural components of automobiles, casings for electronics, etc. owing to their low density, good castability and high specific strength \citep{Mordike2001}. Moreover, they are considered for biodegradable, non-permanent implants in orthopaedic applications owing to their biocompatibility \citep{Echeverry-Rendon2019, LDK21}. Magnesium has a HCP crystal structure, and can accommodate plastic deformation by dislocation slip along different slip planes (basal, prismatic slip and pyramidal) as well as by twinning. The $\{1\bar{1}02\}<\bar{1}101>$ tension twin is the most typical in Mg alloys and is associated with a lattice rotation of $86^{\circ }$ around $[11\bar{2}0]$ together with a shear strain $\gamma =0.129$ on the $\{1\bar{1}02\}$ plane along the $<\bar{1}101>$ direction. While plastic deformation by dislocation slip can be accurately simulated by means of crystal plasticity (CP) in combination with finite elements or fast Fourier transform (FFT) algorithms \citep{Segurado2018}, similar degree of success has not been achieved in the case of twinning. In particular, consideration of twin caused lattice rotation and eigen strain as well as twin produced new interface on constitutive models are not easy tasks.

The first CP framework including the different slip systems in Mg plus twinnin g as a pseudo slip system was solved using the visco-plastic self-consistent method \citep{Lebensohn1993, Proust2007, Beyerlein2008}. This tool was able to predict the tension-compression asymmetry in the stress-strain curves of textured Mg polycrystalline alloys as well as the texture evolution \citep{Lebensohn1993, Proust2007, Beyerlein2008}. Afterwards, the mechanical response of Mg polycrystals was analyzed by mean of the finite element simulation of a representative volume element of the microstructure and the crystal plasticity constitutive model accounted for dislocation slip and twinning as an additional pseudo slip system \citep{Kalidindi1998}. This approach provides an accurate and physically-based representation of this phenomenon at the microscopic level within each grain \citep{Roters20101152} and has been widely used in the literature to simulate the deformation of Mg alloys \citep{AZ31_Kalidindi, AZ31_Joshi, Herrera-Solaz2014a, HHP14} including the effect of de-twinning under reversed deformation \citep{Briffod2019, Zhang2021}. Nevertheless, this approach cannot take into account features associated with the presence of a twinned region in the crystal, namely the twin morphology, the stress concentrations near the twin tip, twin-twin junctions and twin/grain boundary interactions where fatigue cracks can nucleate \citep{Jamali2022}.

Phase-field (PF) models are well suited to analyze problems in which the evolution of the boundaries is precisely part of the solution, such as twin propagation \citep{Clayton2011,clayton2016phase, pi2016phase, liu2018formation}. Mechanical twin PF models  often use non-conserved phase field theory which only includes order parameters and their gradients instead of conserved PF models which consider order parameters and chemical composition at the same time. The local Gibbs free energy density is formulated as function of stress, strain, temperature, order parameters and their spatial gradients \citep{Cahn1958, Allen1979, Clayton2011, Kondo2014, Kondo2015} and the system evolves towards a state with minimum free energy subject to the boundary conditions. As the phase change is completely driven by the free energy, there is no need to make assumptions about whether twinning or de-twinning will take place. In addition, there is no need to track the location of the interface between different phases and multiple twin variants -- or even other mechanisms such as fracture -- can be accounted in one simulation framework \citep{Clayton2011}. 

Obviously, the coupling of CP and PF is an attractive approach to simulate the plastic deformation of Mg alloys including rigorously the effect of dislocation slip and twinning  \citep{Kondo2014, liu2018integrated, zhao2020finite, Ghosh2018, Vidyasagar2018, Hu2021} but it also faces some challenges. The order parameters take values of zero or one far away from the interfaces and there are well-defined stiffness matrix and slip systems. However, the order parameters enable interpolation between different pure phases inside phase boundary regions and the material is a mixture of different phases with different volume fractions in which it is difficult to define the stiffness matrix, slip systems, flow and hardening laws. Furthermore, twin propagation takes place at high velocity (as compared with dislocation slip) and very small time steps are necessary to achieve convergence in addition to a very fine discretization to represent the twin/matrix interface region. This combination leads to huge computational costs and require the implementation of sophisticated multi-time-domain time algorithms which use different time steps for the bulk and twin domains \citep{Ghosh2018}.

The current paper presents a simulation framework which combines the CP model (solved using the finite element method) and the PF model (solved using  the FFT algorithm). As in \cite{Kalidindi1998}, basal slip systems, prismatic and pyramidal slip systems were considered in matrix material. However, \cite{Kalidindi1998} treated twins as pseudo slip systems which need flow and hardening laws, while the phase field model was used in this approach to account for  strain accommodation by twinning  and it was not necessary to formulate hardening and flow rules of twinning in the CP model. As another extension, plastic deformation on basal slip systems, prismatic and pyramidal slip in twinned material with rotated Schmid tensors was calculated explicitly. The multi-phase polynormal term and multi-phase gradient terms of \citep {Steinbach1996} were kept, while the mechanical part of phase field potential was derived based on finite deformation framework. As the multiplicative decomposition was adopted, the current model could be used to investigate problems involving large deformations. Furthermore, it is possible to push back the phase field evolution law from intermediate configuration to the reference configuration. The advantage of doing this is using a fixed FFT grid to solve the phase field evolution. 

The governing equation of linear moment balance was solved using Abaqus finite element code using a UMAT subroutine while the non-linear higher order partial differential equations of PF model were solved by an iterated FFT algorithm. Compared with previous CP-PF strategies, the current model uses different discretizations of the simulation domain to solve the PF and the CP equations but with the same time step. As a result, the simulation effort is significantly reduced and provides good results because the FFT cell is much finer than the FEM discretization. Moreover, it can be implemented in standard software platforms and because it is based on the multi-phase field framework it can be used to study twin-twin interactions, twin-grain boundary interactions as well as twin-precipitate interactions at the micrometer length scale.

This strategy is used to simulate the propagation of a twin in a single crystal Mg micro-pillar deformed in compression along the $[10\bar{1}0]$ direction \citep{Wang2020}, as well as twining inside a polycrystal. The paper proceeds as follows. Section 2 presents the CP model for slip and PF model for twinning. Section 3 introduces the iterated FFT algorithm while the application of the strategy to simulate the compression of the micro-pillar is detailed in section 4, the polycrystal simulation is given in section 5. Important discussion of mesh size influence, numerical efficiency as well as FEM-FFT mapping strategies were reported in section 6. The main conclusions of the paper are summarized in section 7. 

\section{CP-PF simulation strategy}

\subsection{Crystal plasticity model}

The CP-PF model includes the main active slip systems in Mg and Mg alloys:  3 basal slip systems $(0001) \left\langle 11\bar{2}0 \right\rangle$, 3 prismatic slip systems $\left\{ 1\bar{1}00 \right\} \left\langle 11\bar{2}0\right\rangle$, and 12 first order pyramidal slip systems $\left\{ 10\bar{1}1 \right\} \left\langle 1\bar{2}13 \right\rangle$, as well as the six variants of tensile twins $\left\{ 10\bar{1}2 \right\} \left\langle \bar{1}011 \right\rangle$. It was assumed that $c/a$ ratio of the HCP lattice was 1.633. Additionally, $\mathbf{Q}_{\alpha}$ is the orientation matrix of the twin
variant $\alpha$.

\subsubsection{Kinematics}

The CP model follows the standard multiplicative decomposition of the deformation gradient into an elastic part and a plastic part  \citep{Lee1967}, 
\begin{equation}
\mathbf{F}=\mathbf{F}_{\text{e}}\mathbf{F}_{\text{p}}.  
\label{Mdecompoition}
\end{equation}

The elastic behavior, plastic deformation by dislocation slip as well as the PF equations for twinning deformation are expressed in the intermediate configuration, given by $\mathbf{F}_{\text{p}}$, as indicated below.   

\subsubsection{Elasticity}

The elastic right Cauchy-Green strain tensor $\mathbf{C}_{\text{e}}$  in the intermediate configuration is given by \citep{Kalidindi1992} 

\begin{equation}
\widetilde{\mathbf{C}}_{\text{e}} =\mathbf{F}_{\text{e}}^{\text{T}}\mathbf{F}%
_{\text{e}} = \widetilde{\mathbf{C}}_{\text{e}}^{\prime}-(\widetilde{\mathbf{%
C}}_{\text{e}}^{\prime}\mathbf{L}_{\text{p}}+\mathbf{L} _{\text{p}}^{\text{T}%
}\widetilde{\mathbf{C}}_{\text{e}}^{\prime}) \Delta t  \label{CEtensor}
\end{equation}

\begin{equation}
\widetilde{\mathbf{C}}_{\text{e}}^{\prime}=\left(\mathbf{F}\mathbf{F}_{\text{%
p0}}^{-1}\right)^{\text{T}} \left(\mathbf{F}\mathbf{F}_{\text{p0}%
}^{-1}\right).  \label{CEmaxtensor}
\end{equation}

\noindent where $\mathbf{F}$ and $\mathbf{F}_{\text{e}}$ stand for the total deformation
gradient and the elastic deformation gradient, respectively, at the current time step. Additionally, $\mathbf{F}_{\text{p0}}$ is the plastic deformation gradient of last time step.

With the help of equations \eqref{CEtensor} and \eqref{CEmaxtensor}, the second Piola Kirchhoff stress $\widetilde{\mathbf{T}}$ in the
intermediate configuration is expressed as

\begin{equation}
\widetilde{\mathbf{T}} = \widetilde{\mathbb{C}} \left[ \frac{1}{2} \left(%
\widetilde{\mathbf{C}}_{\text{e}}^{\prime}-(\widetilde{\mathbf{C}}_{\text{e}%
}^{\prime}\mathbf{L}_{\text{p}}+\mathbf{L}_{\text{p}}^{\text{T}}\widetilde{%
\mathbf{C}}_{\text{e}}^{\prime}) \Delta t -\mathbf{I} \right) \right].
\label{PK2i}
\end{equation}

As the material point is a mixture of matrix with volume
fraction $f_{\text{M}}=\left(1-\overset{N_{\text{tw}}}{\underset{\alpha =1}{%
\sum}}f_{\alpha}\right)$ and twin with the volume fraction $1-f_{\text{M}}$, an average elastic stiffness is proposed according to

\begin{equation}
\widetilde{\mathbb{C}}=f_{\text{M}}\widetilde{\mathbb{C}}_{\text{matrix}}+%
\overset{N_{\text{tw}}}{\underset{\alpha =1}{\sum}}f_{\alpha}\widetilde{%
\mathbb{C}}_{\text{twin}}^{\alpha}  \label{stiffM}
\end{equation}

\noindent where $\widetilde{\mathbb{C}}_{\text{matrix}}$ is the stiffness of the matrix and 
$\mathbb{C}_{\text{twin}}^{\alpha}$ is the stiffness of twin variant, i.e. $(\mathbb{C}_{\text{twin}}^{\alpha})_{ijkl}=(\widetilde{\mathbb{C}}_{\text{matrix}})_{abcd}(\mathbf{Q}_{\alpha})_{ia}(\mathbf{Q}_{\alpha})_{jb}(\mathbf{Q}_{\alpha})_{kc}(\mathbf{Q}_{\alpha})_{ld}$. Thus, the twin volume fraction $f_{\alpha}$ will influences the elastic behavior as well as the strain energy which will be used in the PF model.

The components of stiffness tensor of matix are listed in  Table \ref{elastic_constants}. 

\subsubsection{Plasticity}

The plastic deformation is accommodated by dislocation slip in matrix, re-slip in the twins
as well as the eigenstrain associated with twinning. Although the interaction among the deformation mechanisms is very complex, the classical CP approach in the intermediate configuration can be used to express the plastic velocity gradient $\mathbf{L}_{\text{p}}$ as \citep{Kalidindi1992, Kalidindi1998},

\begin{equation}
\mathbf{L}_{\text{p}}= f_{\text{M}}\overset{N_{\text{sl}}}{\underset{\alpha=1%
}{\sum}}\dot{\gamma}_{\alpha}\widetilde{\mathbf{M}}_{\alpha} + \overset{N_{\text{tw}}}{\underset{\beta=1}{\sum}} 
\overset{N_{\text{sl}}}{\underset{\alpha=1}{\sum}} f_{\beta} \dot{\gamma}%
_{\alpha} \mathbf{Q}_{\beta}\widetilde{\mathbf{M}}_{\alpha}\mathbf{Q}_{\beta}^{\text{T}} +\gamma_{\text{tw}} 
\overset{N_{\text{tw}}}{\underset{\beta=1}{\sum}}\dot{f}_{\beta} \widetilde{\mathbf{M}}^{\prime}_{\beta}
\end{equation}

\noindent where $\widetilde{\mathbf{M}}=\widetilde{\mathbf{s}}\otimes\widetilde{\mathbf{m}}$ and $\widetilde{\mathbf{M}}^{\prime}=\widetilde{\mathbf{s%
}}^{\prime}\otimes\widetilde{\mathbf{m}}^{\prime}$ stand for the Schmid tensors of the
slip system and of the  twin system, respectively.  $\gamma_{\text{tw}}=0.129$
represents the eigenstrain associated with the tensile twin and $\dot{f}_{\beta}$ is the rate of evolution of the twin volume
fraction. 

Once the velocity gradient $\mathbf{L}_{\text{p}}$ is known the current plastic deformation
gradient can be obtained as: 
\begin{equation}
\mathbf{F}_{\text{p}}=\mathbf{(I}+\mathbf{L}_{\text{p}} \Delta t)\mathbf{F}_{%
\text{p0}}  \label{FpUpdate2}
\end{equation}

\noindent where $\mathbf{I}$ is the second order identity tensor and provided that the time step $\Delta t$ is small enough.

\subsubsection{Flow and hardening laws}

The resolved shear stress, $ \tau_{\alpha}$, which is the projection of the
stress tensor to the Schmid tensor,   is expressed in a slip system of the matrix as

\begin{equation}
\tau_{\alpha} = \widetilde{\mathbf{T}} \cdot \widetilde{\mathbf{M}}_{\alpha}  
\label{tauSlip}
\end{equation}

\noindent and in a slip system of the twin variant $\beta$ as

\begin{equation}
\tau_{\alpha} = \widetilde{\mathbf{T}} \cdot ( \mathbf{Q}_{\beta} \widetilde{\mathbf{M}}_{\alpha} \mathbf{Q}_{\beta}^{\text{T}
} ).  
\label{tauSlipTwin}
\end{equation}

The shear rate is given by 

\begin{equation}
\dot{\gamma}_{\alpha}=\dot{\gamma}_{0} \left| \frac{\tau_{\alpha}}{\hat{\tau}%
_{\alpha}} \right|^{\frac{1}{m}} sign(\tau_{\alpha})  \label{dgmdt_slip}
\end{equation}

\noindent where $\dot{\gamma}_{0}$ is the reference shear rate parameter and $m$ the
strain rate sensitivity. The variable $\hat{\tau}_{\alpha}$ represents the
isotropic strain hardening 

\begin{equation}
\dot{\hat{\tau}}_{\alpha}=\underset{\beta=1}{\overset{N_{\text{sl}}}{\sum}}%
h_{\alpha \beta} H_{\beta} (1-\frac{\hat{\tau}_{\beta}}{\hat{\tau}^{s}})^{a_{%
\text{ss}}} |\dot{\gamma}_{\beta}|  \label{dtaucdt_slip}
\end{equation}

\noindent where $h_{\alpha \beta}$ was the latent hardening parameter between slip
systems $\alpha$ and $\beta$. 

The flow and hardening laws of slip in matrix and of re-slip in the twin variants
are equivalent,  and the only difference is that the Schmid tensor $\widetilde{\mathbf{M}}_{\alpha}$ and the rotated Schmid tensor 
$\mathbf{Q}_{\beta}\widetilde{\mathbf{M}}_{\alpha}\mathbf{Q}_{\beta}^{\text{T}}$ are used to calculate the 
resolved shear stress, respectively.

\subsection{Phase field model of twinning}

According to the PF concept for multiphase materials \citep {Steinbach1996}, the free energy functional of the matrix-twin
microstructure includes three energy terms: the double well energy which tends to separate different phases, the interfacial energy which depends on the gradient of order parameter and the mechanical strain energy

\begin{equation}
G = G_{\text{dw}} + G_{\text{gd}} + G_{\text{me}}.  \label{totalG}
\end{equation}

The volume fraction of twin $f_{\alpha}$ is the order
parameter and  the double well energy of a multiphase system is expressed as 
 
\begin{equation}
G_{\text{dw}}= c_{1} \sum_{\alpha,\beta; \alpha<\beta}^{N_{\text{tw}}+1}
f_{\alpha}^{2}f_{\beta}^{2}  \label{totalDW}
\end{equation}

\noindent where $c_{1}$ is a model parameter. 

The interfacial energy contribution is given by
 
\begin{equation}
G_{\text{gd}}= c_{2} \sum_{\alpha,\beta; \alpha<\beta}^{N_{\text{tw}}+1}
\left\| f_{\beta} (\boldsymbol{\nabla}f_{\alpha}) - f_{\alpha} (\boldsymbol{%
\nabla} f_{\beta} ) \right\| ^{2}  \label{totalGD}
\end{equation}

\noindent where $c_{2}$ is another model parameter. 

Finally the stored strain energy density is given by

\begin{equation}
G_{\text{me}}=\frac{1}{2} \widetilde{\mathbb{C}}_{ijkl} \widetilde{E}_{ij} 
\widetilde{E}_{kl}  \label{totalME}
\end{equation}%

\noindent where $\widetilde{\mathbf{E}}=\frac{1}{2}(\widetilde{\mathbf{C}}_{\text{e}}-\mathbf{I})$ is the elastic strain of the matrix-twin microstructure in the intermediate configuration. As shown in \eqref{CEmaxtensor}, the averaged stiffness tensor is a linear function of the twin volume fraction. To keep the model simple,  the elastic strain tensor is considered to be constant during the PF simulation.

The evolution of the twin volume fractions has to move the functional $G$ towards its minimum. Thus, rate of twin volume fraction is given by 
\begin{equation}
\dot{f_{\alpha}}=-c_{3} \left( \frac{\delta G}{\delta f_{\alpha}} \right) =
-c_{3} \left(\frac{\partial G}{\partial f_{\alpha}}- \frac{\partial}{%
\partial x_{i}} \frac{\partial G}{\partial \frac{\partial f_{\alpha}}{%
\partial x_{i}}} \right)  \label{fRate1}
\end{equation}%

\noindent and if equations \eqref{stiffM}, \eqref{totalDW}, \eqref{totalGD} and \eqref{totalME}
are substituted into equation \eqref{fRate1}, the evolution of the twin volume fraction is given by  

\begin{equation}
\begin{aligned} 
\dot{f_{\alpha}} = 
& - c_{3}c_{1} \sum_{\beta; \beta \neq \alpha}^{N_{\text{tw}}+1} f_{\alpha}f_{\beta}(f_{\beta}-f_{\alpha}) 
  + c_{3}c_{2} \sum_{\beta; \beta \neq \alpha}^{N_{\text{tw}}+1} f_{\beta} f_{\alpha,ii}  
  - c_{3}c_{2} \sum_{\beta; \beta \neq \alpha}^{N_{\text{tw}}+1} f_{\alpha} f_{\beta,ii}  \\
& - \frac{c_{3}}{2} \left(\widetilde{\mathbb{C}}_{\text{twin}}^{\alpha}\right)_{ijkl}\widetilde{E}_{ij} \widetilde{E}_{kl}  + c_{3} \widetilde{\mathbb{C}}_{ijkl}\widetilde{E}_{ij}
\left(\widetilde{\mathbf{C}}^{\prime}_{\text{e}}\widetilde{\mathbf{M}}_{\alpha}^{\prime} \gamma_{\text{tw}} \right)_{kl}
\end{aligned}  
\label{fRate2}
\end{equation}

The first term on the right hand side of equation \eqref{fRate2} represents the driving force of phase transformation owing to the gradient of the volume fraction itself, the second term stands for gradient of the volume fraction of other twin variants, the third one for the double well potential energy and the last ones for the mechanical strain energy contribution, respectively. 

Different continuum models for twin nucleation have been proposed in the literature. For instance, \cite{Cheng2021} presented amodel  based on non-planar dissociation of a sessile <c+a> pyramidal dislocation, while twin nucleation triggered by the interaction of basal slip bands with grain boundaries or free surfaces has been often reported in experiments \citep{Ventura2021}. In the absence of a well established theory to account for twin nucleation, it was assumed in our model that twins were nucleated in the presence of stress concentrations and  the last two terms of equation \eqref{fRate2} are responsible for introduction of sources terms $r_{\alpha}^{i}$ in equation \eqref{fRate3} for twin nucleation in the presence of stress concentrations. These sources terms will only lead to the nucleation of a twin variant if is energetically favoured.

It should be noted that the volume fractions of twin variants and matrix are bounded by the following conservation law 

\begin{equation}
\sum_{\alpha}^{N_{\text{tw}}+1}f_{\alpha}=1. 
\label{conserveLaw}
\end{equation}

\section{Numerical implementation}
\label{VoronoiMapMethod}

The coupled CP and PF equations were solved using different discretizations of the simulation domain using the same time step for both of them. The linear momentum balance equations for the CP model were solved using the Abaqus finite element model with a UMAT user subroutine whose details were reported in \cite{Zhang2021}. The PF equations were solved with a FFT algorithm using a regular grid and the code was embedded into the UMAT subroutine with the help of UEXTERNALDB subroutine of Abaqus. The FFT cell was finer than the finite element discretization and it was extended beyond the real finite element simulation domain to deal with possible complicated sample shapes. Thus, the actual domain was extended by adding a pad layer for the FFT domain in which the mobility of the PF model was very low.

The numerical implementation at each time step always began with the CP finite element simulation. At the end the time step the phase field source rate $\dot{f_{\alpha}}$ defined in equation \eqref{fRate2} of each Gauss point was calculated. To this end, the Gauss points were taken as seeds of a Voronoi tessellation and the FFT domain was tessellated. Each FFT cell is found within a tessellation domain and the information associated with this Gauss point was transferred to the FFT cell. This approach is more efficient from the numerical viewpoint than interpolations techniques and provides good prediction capability. Comparison about different mapping approaches were reported in section \ref{DisSizeMap}.

The control equation of twin volume fraction, equation \eqref{fRate2}, is non-linear and cannot not be solved directly. Thus, all nonlinear parts of twin volume fraction rate were formulated inside the source term $r_{\alpha}^{i}$ using the variables from the iteration number $i-1$. Thus, equation \eqref{fRate2} has been linearized in each time step according to

\begin{equation}
\frac{f_{\alpha}^{i}}{\Delta t} - c_{3}c_{2} \left[\sum_{\beta; \beta \neq
\alpha}^{N_{\text{tw}}+1} f_{\beta}^{i-1} \right] f_{\alpha,jj}^{i} +
r_{\alpha}^{i} = 0  
\label{fRate3}
\end{equation}

\noindent and

\begin{equation}
\begin{aligned} 
r_{\alpha}^{i} 
& = -\frac{f_{0\alpha}}{\Delta t} 
    + c_{3}c_{1} \sum_{\beta; \beta \neq \alpha}^{N_{\text{tw}}+1} f_{\alpha}^{i-1}f_{\beta}^{i-1}(f_{\beta}^{i-1}-f_{\alpha}^{i-1})
    + c_{3}c_{2} \sum_{\beta; \beta \neq \alpha}^{N_{\text{tw}}+1} f_{\alpha}^{i-1} f_{\beta,jj}^{i-1} \\ 
& + \frac{c_{3}}{2} \left(\widetilde{\mathbb{C}}_{\text{twin}} ^{\alpha}\right)_{ijkl}\widetilde{E}_{ij} \widetilde{E}_{kl} - c_{3} \widetilde{\mathbb{C}}_{ijkl}\widetilde{E}_{ij}
\left(\widetilde{\mathbf{C}}^{\prime}_{\text{e}}\widetilde{\mathbf{M}}_{\alpha}^{\prime} \gamma_{\text{tw}} \right)_{kl}
\end{aligned}  
\label{fRate4}
\end{equation}
\noindent where $i$ in $f_{\alpha}^{i}$ stands for the iteration number of the FFT simulation. 

Equations \eqref{fRate3} and \eqref{fRate4} are linear functions of $f_{\alpha}^{i}$ and they can be efficiently solved in the frequency domain using the FFT method. The iterative strategy to solve the twin volume fraction evolution is detailed in Table \ref{Algorithm_PF} and will be finished when the convergence criteria are satisfied. The algorithm includes four tasks which are related to the indexes $(k, j, \alpha, i)$ to solve the multi-variable phase field evolution equations. The first task is to divide the whole FFT domain into independent sub-domains ($k$). The second task is to ensure the conservation law of phase field values in each FFT cell by an iteration ($j$), taking into account the contribution of the different twin variants (denoted by $\alpha$). Finally, the phase field governing equations of each twin variant have to be iteratively solved ($i$). The twin volume fractions resulting from the PF model are transferred to the Gauss points of the finite element discretization. 
  
\begin{table}[!tbp]
\centering%
\begin{tabular}{l|l}
\hline
\hline
{\calligra 1}  & calculate phase field source rate $r_{\alpha}^{i}$ of equation \eqref{fRate4} at Gauss points \\ 
{\calligra 2}  & map $r_{\alpha}^{i}$ from FEM domain to FFT domain \\
\hline
{\calligra 3}  & repeat FFT: in domain $k$ volume fraction $f$ is a variable, in other domains $f=f_{0}$ \\
{\calligra 4}  & \ \ \ \ check conservation equation \eqref{conserveLaw} with iteration $j$ \\
{\calligra 5}  & \ \ \ \ \ \ \ \ loop for twin systems $\alpha$ \\
{\calligra 6}  & \ \ \ \ \ \ \ \ \ \ \ \ twin volume fraction iteration $i$ \\
{\calligra 7}  & \ \ \ \ \ \ \ \ \ \ \ \ \ \ \ \ guess twin volume fraction \\
{\calligra 8}  & \ \ \ \ \ \ \ \ \ \ \ \ \ \ \ \ solve equation (\ref{fRate3}) in FFT space \\
{\calligra 9} & \ \ \ \ \ \ \ \ \ \ \ \ \ \ \ \ if residual $ \left|  f_{\alpha}^{i}-f_{\alpha}^{i-1} \right| \leq ERR_{\text{f}}$, goto {\calligra 10} \\
{\calligra 10} & \ \ \ \ \ \ \ \ calculate volume fraction summation \\ 
{\calligra 11} & \ \ \ \ \ \ \ \ if residual $ \left|  \sum{f_{\alpha}^{i}} -1 \right| \leq ERR_{\text{f}}$, goto {\calligra 12} \\ 
\hline
{\calligra 12} & map volume fraction $f$ from FFT cells to FE Gauss points \\ 
\hline
\hline
\end{tabular}%
\caption{Iterative algorithm of polycrystal multi-variable phase field model.}
\label{Algorithm_PF}
\end{table}

\section{Application to simulate micro-pillar compression tests}

\subsection{Experimental evidence}

The coupled CP-PF model presented above has been used to simulate the compression of a single crystal micro-pillar of Mg and to compare with the experimental data reported in \cite{Wang2020}, which are detailed below for the sake of completion. Micro-pillars with square cross-section of 5.6 $\times$ 5.6 $\mu$m$^2$ and 13.1 $\mu$m in length were milled using focus ion beam in a pure Mg sample. The micro-pillars were subjected to compression along the longest axis that was parallel to the $[10\bar{1}0]$ direction of the Mg HCP lattice using a nanoindenter equipped with a diamond flat punch of 15 $\mu$m in diameter at an at a constant strain rate of 10$^{-3}$ s$^{-1}$ under the displacement control up to a maximum strain of 10\%. The load-displacement curves were corrected by applying the Sneddon method to account for the compliance associated with the elastic deflection of the base of the micro-pillar. The engineering stress-strain curves were obtained from the corrected curves from the cross-sectional area and the height of the micro-pillar before deformation \citep{Wang2020}. 

\begin{figure}[!tbp]
\centering
\includegraphics[width=0.5\textwidth]{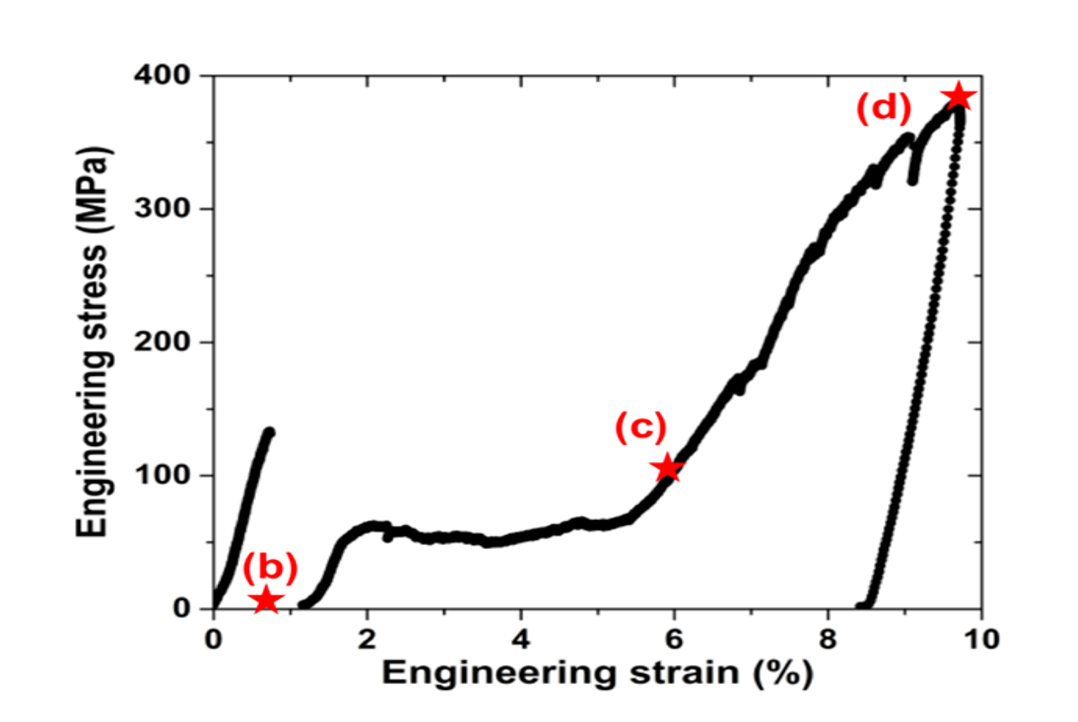}
\caption{Experimental stress-strain curve of a Mg micro-pillar deformed in compression parallel to the $[10\bar{1}0] $  direction of the Mg HCP lattice. Adapted from paper \citep{Wang2020}.}
\label{fig:expSE}
\end{figure}

\begin{figure}[tbp]
\centering
    \includegraphics[width=0.8\textwidth]{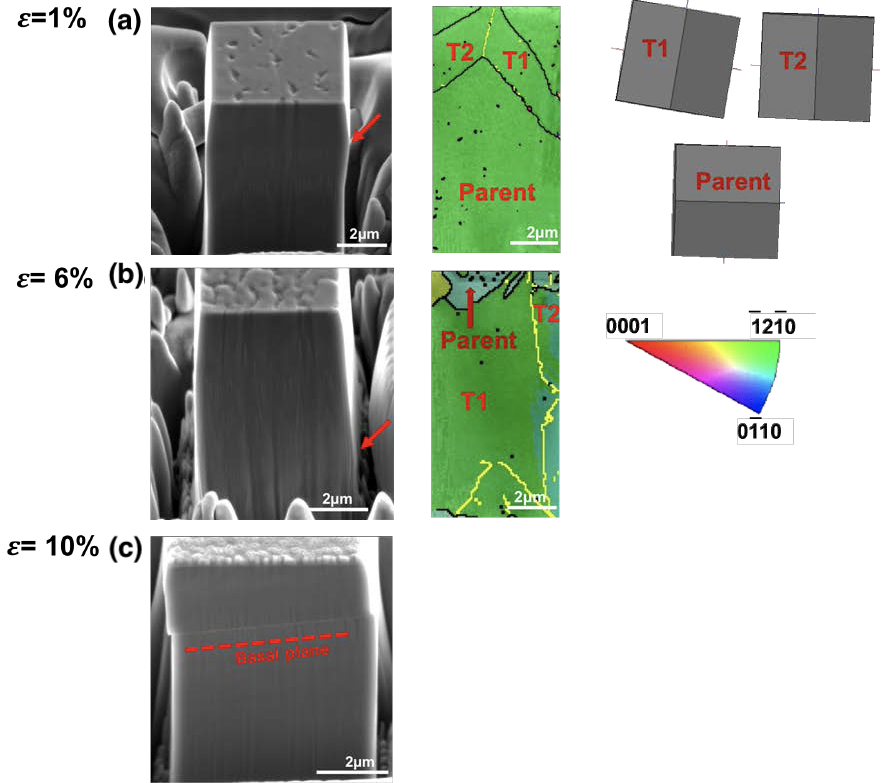}
\caption{(a) Secondary electron micrographs of the deformed micropillar after the load drop at 1\% strain. (b) Idem at 6\% strain. (d) Idem at 10\% strain. The orientation maps of a thin lamella extracted from the deformed micropillars in (a-b) are shown to the right of the corresponding micrographs. They show the parent grain and the tensile twin variants T1 and T2. Black and yellow lines stand for the boundary between the twin variants and the parent grain and the boundary between the two twin variants, respectively. The orientation of the parent grain and of the tensile twin variants are shown in the schematics to the right. Adapted from \citep{Wang2020}.}
\label{fig:expDetail}
\end{figure}

Several micro-pillars of pure Mg were tested and one representative stress-strain curve is plotted in Figure \ref{fig:expSE}. The initial region of the stress-strain curve was linear up to the compressive stress reached $\approx$ 140 MPa. A sudden strain burst occurred at this point and the load dropped to zero as the flat punch lost contact with the micro-pillar surface. Further deformation of the micro-pillar was associated with a short elastic region followed by a plateau at $\approx$ 60 MPa and by another region with a strong strain hardening. To understand the dominant deformation mechanisms, different tests were stopped at the points marked with (b), (c) and (d) in Figure \ref{fig:expSE} and observed in the scanning electron microscope. Moreover, lamella of the cross-sections of the deformed micro-pillars were extracted using the standard lift-out technique and observed in the transmission electron microscope. These analysis showed that the strain burst marked with (b) in Figure \ref{fig:expSE} was associated with the nucleation of two $\left\{10\bar{1}2\right\}$ tension twin variants at the top of the micro-pillar, marked with T1 and T2 in Figure \ref{fig:expDetail}(a). The  $(01\bar 12)[0\bar 111]$ twin variant $T_1$ with the highest Schmid factor was  grew along the micro-pillar at constant stress and the micro-pillar was fully twinned at (c), as it is shown in Figure \ref{fig:expDetail}(b). Finally, the micro-pillar deformed up to (d) showed traces of basal slip on the lateral surface of the micro-pillar, Figure \ref{fig:expDetail}(b), although the orientation within the twinned region is not favourably oriented for basal slip ($\approx$ 8$^\circ$ misorientation with respect to the $c$-axis). Additionally, the analysis of the lamella by transmission electron microscope showed evidence of $\left\langle c+a\right\rangle $ pyramidal dislocations that were responsible for the large strain hardening. More details can be found in \cite{Wang2020}.

\subsection{Simulation model}

\begin{figure}[!tbp]
\centering
\includegraphics[width=0.9\textwidth]{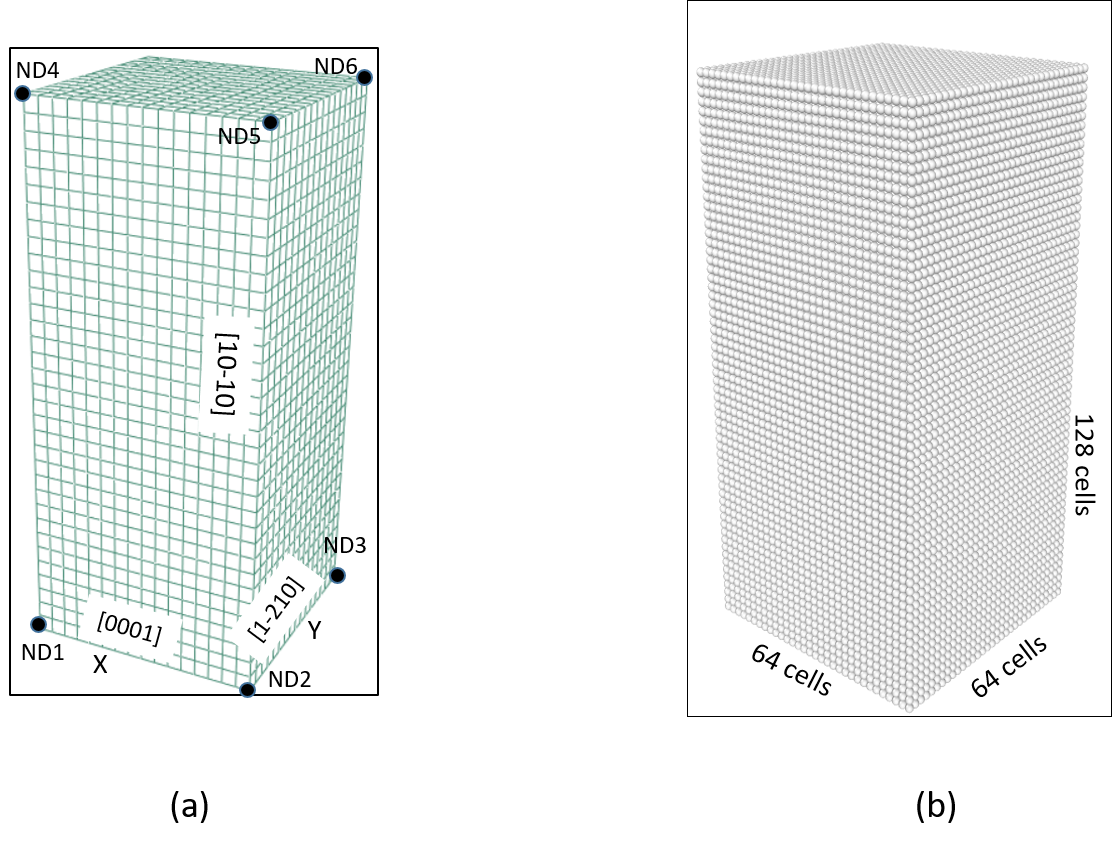}
\caption{(a) Finite element discretization and (b) FFT cells of the micro-pillar. The FFT domain includes a pad layer outside of the micro-pillar to take care of the periodic boundary conditions.}
\label{fig:elements}
\end{figure}

The micro-pillar model used in the simulations is depicted in Figure \ref{fig:elements}. The micro-pillar was oriented with the $[10\bar{1}0]$ direction parallel to the loading axis and the lateral faces of the micro-pillar were parallel to $(0001)$ and  $(1\bar{2}10)$. The dimensions of the model micro-pillar were identical to the experimental one. It was discretized with 8704 elements (C3D8R of Abaqus), Figure \ref{fig:elements}(a). The FFT cell discretization was much finer with $64 \times 64 \times 128$ cells and it was extended outside of the micro-pillar. 
The thickness of the pad layer in the FFT discretization outside of the micro-pillar was approximately 10\% of the width of the micro-pillar (0.5 $\mu$m) in all directions, Figure \ref{fig:elements}(b). Periodic boundary conditions were adopted in the larger FFT domain which contains the real material region and the pad region. Phase field flux from the real material region to the pad region is suppressed as we assigned very low mobility to the pad cells. As there is not flux from the real material to the pad region, a Neumann boundary condition was adopted. The phase field values of FFT domain in the pad region (phase field value, source rate and constants such as $c_{3}$ and $c_{2}$) were determined according to Voronoi tessellation principle, e.g. the source rate was calculated with equation (\ref{fRate4}).

The boundary conditions of the finite element domain are the following. The vertical displacements of the nodes at the bottom surface of the micro-pillar ($Z$ =0) were constrained. Load was introduced by increasing the vertical displacement of all the nodes on the upper surface of the micro-pillar. The velocity of these nodes was set to achieve a strain rate of 10$^{-3}$ s$^{-1}$. In order to avoid compression-induced bending, the degrees of freedom along $\mathbf{X}$ and $\mathbf{Y}$ in ND2 and ND5, along $\mathbf{X}$ in ND3 and ND6, as well as along $\mathbf{Y}$ in ND1 and ND4, were also constrained.
Periodic boundary conditions along $\mathbf{X}$, $\mathbf{Y}$ and $\mathbf{Z}$ were used in the phase field simulation.

\subsection{Parameter identification}

The elastic constants for pure Mg are depicted in Table \ref{elastic_constants} \citep{Zhang2021}. The parameters that control the plastic deformation along basal, prismatic and pyramidal slip planes were also obtained from \cite{Zhang2021} and \cite{Wang2020} and they can be found in Table \ref{CP_paramter_part1} and Table \ref{CP_paramter_part2}. the first one includes the reference shear rate, $\dot\gamma_0$, the strain rate sensitivity exponent $m$, the slip-slip hardening exponent $a_{ss}$ and the latent hardening parameters $h_{\alpha\beta}$ between coplanar and non-coplanar slip systems. The second one summarizes the initial CRSS, $g^{ini}$, the saturated CRSS, $g^{sat}$, and the hardening modulus, $h_0$ for the three different slip systems. 

\begin{table}[!tbp]
\centering%
\begin{tabular}{|c|c|c|c|c|}
\hline
$C_{11}$ & $C_{12}$ & $C_{33}$ & $C_{13}$ & $C_{44}$ \\ \hline
59.4 & 25.6 & 61.6 & 21.4 & 16.4 \  \\ \hline
\end{tabular}%
\caption{Elastic constants of pure Mg single crystal \citep{Zhang2021} in GPa.}
\label{elastic_constants}
\end{table}

\begin{table}[!tbp]
\centering%
\begin{tabular}{|l|l|l|}
\hline
$\dot{\gamma}_{0}$ & 0.001 1/s   \\ 
$m$ & 0.1  \\ 
$a_{\text{ss}} $ & 1.1 \\ 
$h_{\alpha \beta}$ (coplanar) & 1 \\ 
$h_{\alpha \beta}$ (non-coplanar) & 1.4 \\
\hline
\end{tabular}%
\caption{Parameters for the flow and hardening laws of the CP model \citep{Zhang2021}.}
\label{CP_paramter_part1}
\end{table}

\begin{table}[!tbp]
\centering%
\begin{tabular}{|l|l|l|l|l|}
\hline
 & basal slip & prismatic slip & pyramidal slip  \\ \hline
$g^{ini}$ & 4.0 MPa & 39 MPa & 75 MPa  \\ \hline
$g^{sat}$ & 4.5 MPa & 55 MPa & 179 MPa \\ \hline
$h_{0}$ & 20 MPa & 150 MPa & 1450 MPa  \\ \hline
\end{tabular}%
\caption{Initial CRSS, $g^{ini}$, saturated CRSS, $g^{sat}$, and hardening modulus, $h_0$ for the three different slip systems in pure Mg.}
\label{CP_paramter_part2}
\end{table}

Regarding the PF model, the double well potential parameter $c_{1}$, the gradient energy parameter $c_{2}$ as well as the phase boundary mobility coefficient $c_{3}$ were determined by fitting the experimental and numerical stress-strain curves of the micro-pillar compression tests and they are listed in Table \ref{CP_paramter_part3}. 
The phase boundary mobility coefficient $c_3$ is a key parameter to control the stable simulation time step, which is  0.0005 s for the magnitude of $c_3$ in Table 5. The simulation shown in Figure \ref{fig:seComparison1} took 6.5 CPU hours in a workstation with one processor.

\begin{table}[!tbp]
\centering%
\begin{tabular}{|l|l|l|l|}
\hline

$c_{1}$ & 31.25 MPa  & \\ \hline
$c_{2}$ & 2.81$\times 10^{-7}$ N & $(10\bar{1}2)$ normal direction \\ \hline
$c_{2}$ & 2.81$\times 10^{-6}$ N & $[\bar{1}011]$ shear direction \\ \hline
$c_{3}$ & 12.0 (MPa $\cdot $ s)$^{-1}$ & \\ 
\hline
\end{tabular}%
\caption{Phase field model parameters.}
\label{CP_paramter_part3}
\end{table}

\subsection{Comparison with experiments}

\begin{figure}[tbp]
\centering
    \includegraphics[width=1.2\textwidth]{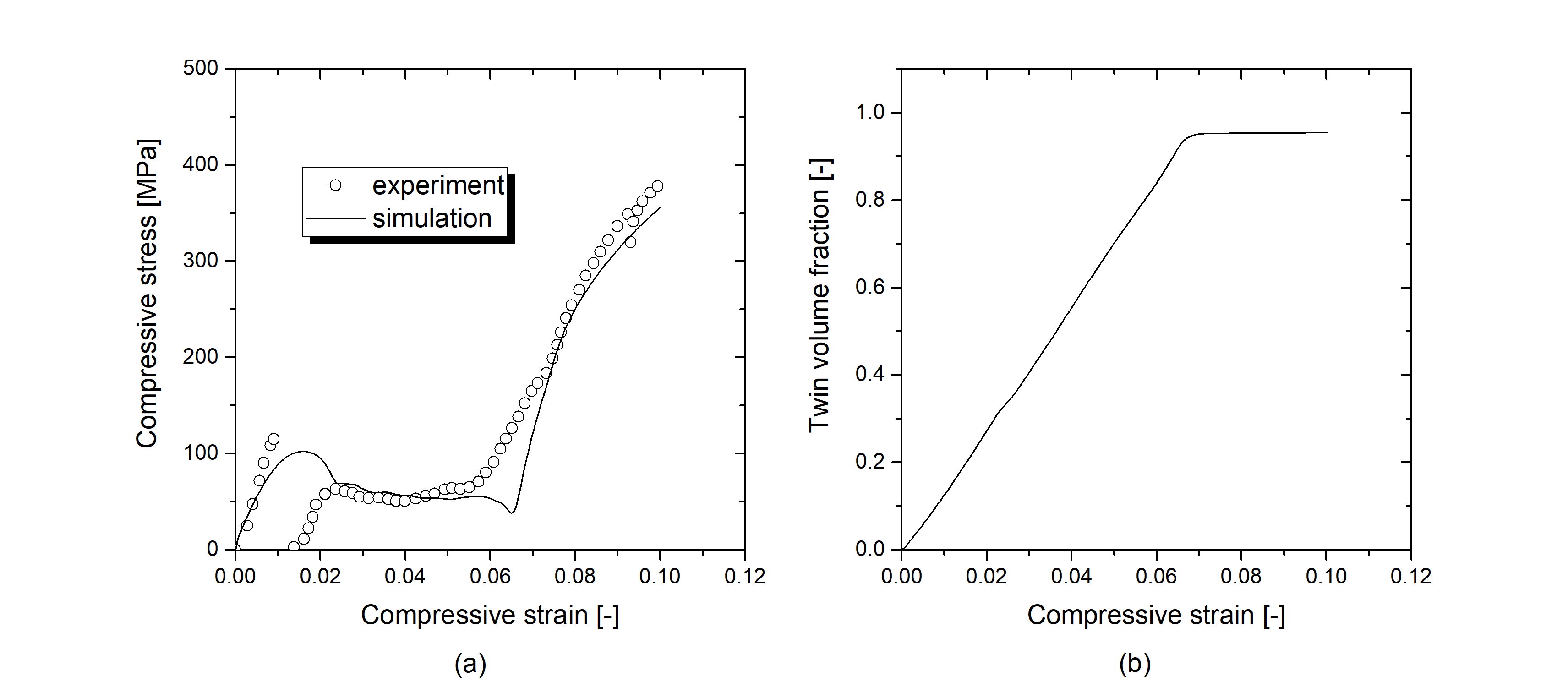}
\caption{Simulated and experimental engineering stress-strain curves of the Mg micro-pillar in compression (a) and evolution of the twin volume fraction with the applied compressive strain according to the simulations (b). The inflection point is at $\epsilon_{33}=0.017$ with peak compression stress before twin nucleation and growth.}
\label{fig:seComparison1}
\end{figure}

\begin{figure}[!tbp]
\centering
\includegraphics[width=0.8\textwidth]{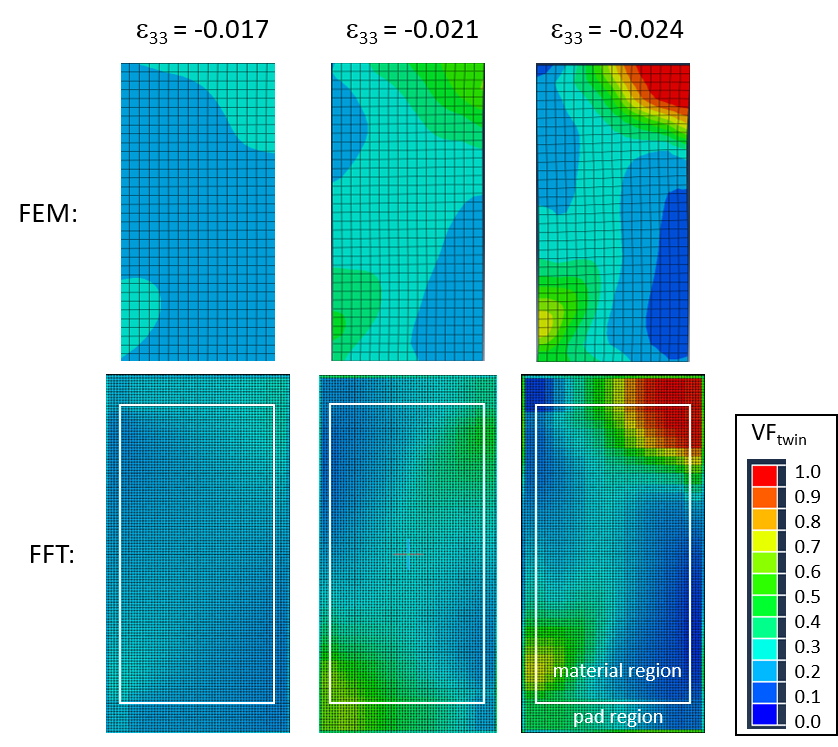}
\caption{Contour plots of volume fraction of twinned material at each Gauss point at the early stages of deformation ($\epsilon_{33}$ = -0.017, -0.021, and -0.024) in the finite element model (FEM) and in the FFT domain showing the process of twin nucleation. The interface between the pad region and the material region in the FFT domain is marked with white lines.}
\label{fig:nucleationDetail}
\end{figure}

\begin{figure}[!tbp]
\centering
\includegraphics[width=1.0\textwidth]{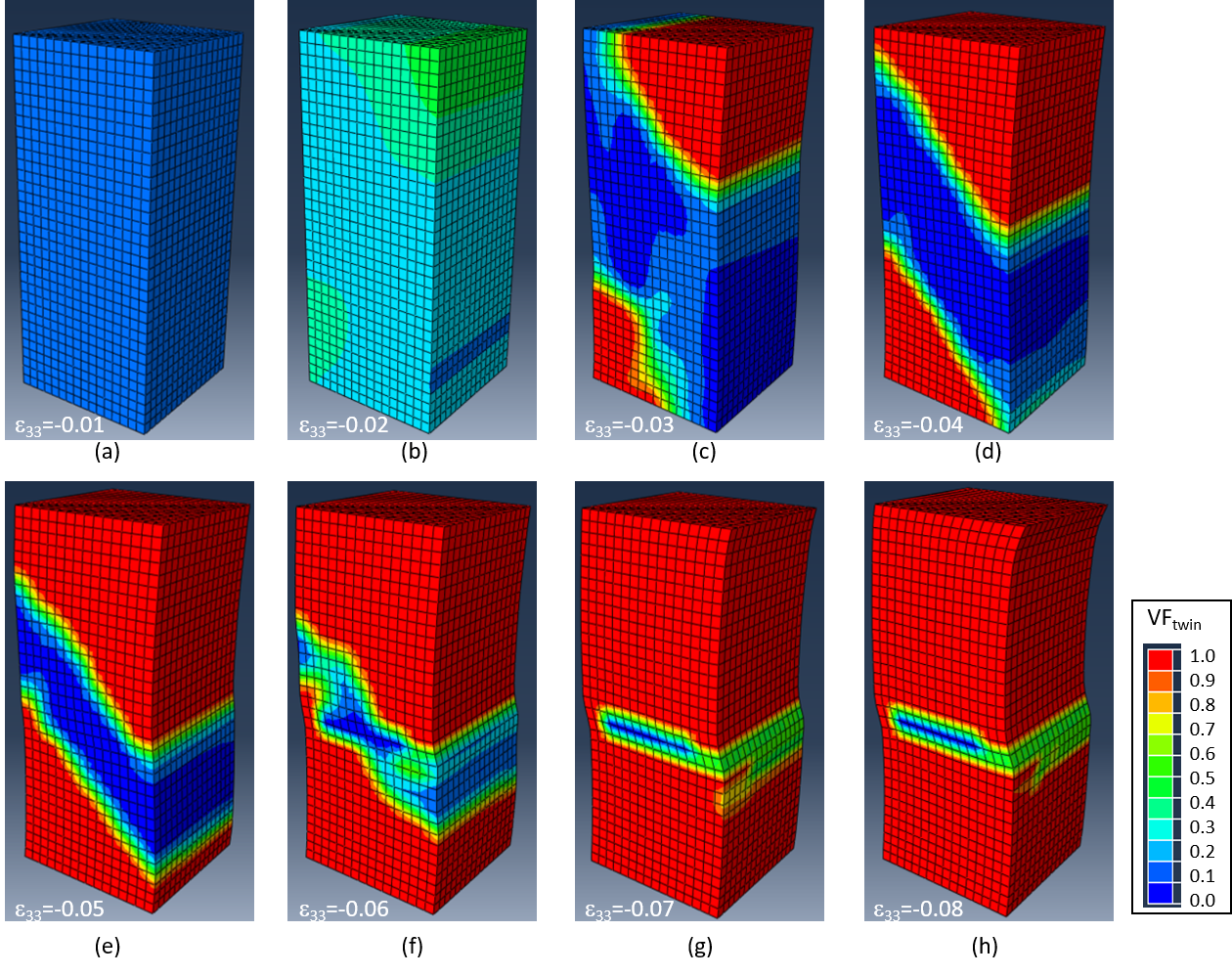}
\caption{Progression of the twinned region as a function of the applied compressive strain. The contour plot stands for the volume fraction of twinned material at each Gauss point.}
\label{fig:twinningStages}
\end{figure}

The simulated stress-strain curve was compared with the experiment in Figure \ref{fig:seComparison1}(a).  The simulations present a short elastic region, followed by a non-linear region  and maximum in the stress as a twin was nucleated at the top of the micropillar. The process of twin nucleation in the simulations is detailed in Figure \ref{fig:nucleationDetail}. In the early stage of micro-pillar compression ($\epsilon$ = -0.017), the phase field variable (volume fraction of twinned material) is slightly different from 0 in the whole pillar due to the stress applied. It increases with the compressive strain but an actual twin variant with a stable phase boundary is only nucleated at the corner on top of the pillar at $\epsilon_{33}=0.024$. After twinning nucleation, the stress carried by the micro-pillar decreased. Moreover, the fast motion of twin-matrix interface caused excessive plastic dissipation  and the stored elastic energy decreased, leading to oscillations in the stress-strain curve. They are followed by a plateau and the twin volume fraction increases linearly with the applied strain in this region. As shown in Figure \ref{fig:seComparison1}(b), the applied strain is mainly accommodated by the eigenstrain associated with the progressive twinning of the micro-pillar. The progression of the twinned region with strain is shown in Figure \ref{fig:twinningStages} where the whole micro-pillar has been twinned at the strain level of $\epsilon \approx $ 6\%. Afterwords, the simulated stress-strain curve shows a step hardening, in agreement with the experimental evidence and the strain hardening rate in the simulations is similar to the experimental results. 

It is interesting to notice the slight tilt of the vertical edge of the micro-pillar at the matrix/twin interface, which is due to the orientation of the (10$\bar 1$2) twin plane with respect to the micro-pillar axis, see Figure \ref{fig:twinningStages}. This tilt propagates downwards along the micro-pillar with the twin/matrix interface, in excellent agreement with the scanning electron microscopy images of Figures \ref{fig:expDetail}(a) and (b). In addition, the simulated twin-matrix interfaces were parallel to $(10\bar{1}2)$ coherent twin plane.

\begin{figure}[!tbp]
\centering
\includegraphics[width=0.8\textwidth]{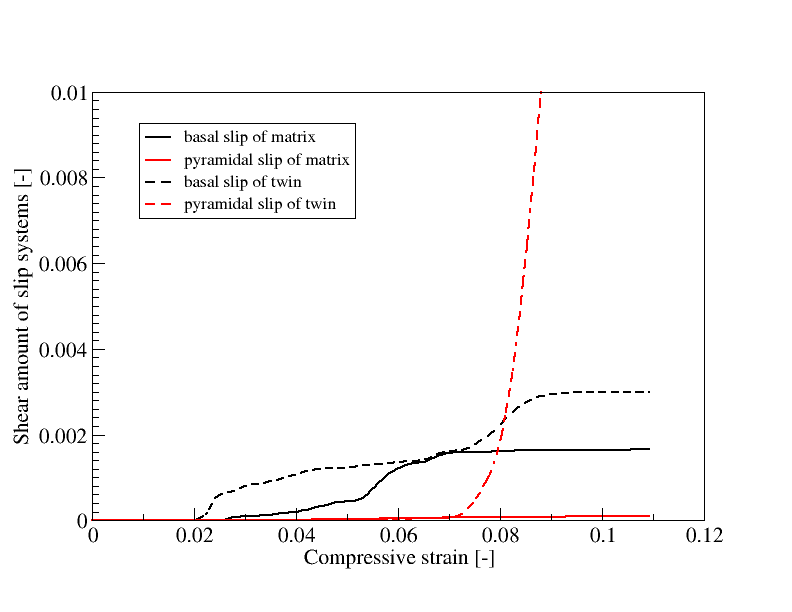}
\caption{Contribution to shear deformation of basal and pyramidal slip systems in the matrix and in the twinned region as a function of the applied strain.}
\label{fig:slipMechanism}
\end{figure}

\begin{figure}[!tbp]
\centering
\includegraphics[width=0.4\textwidth]{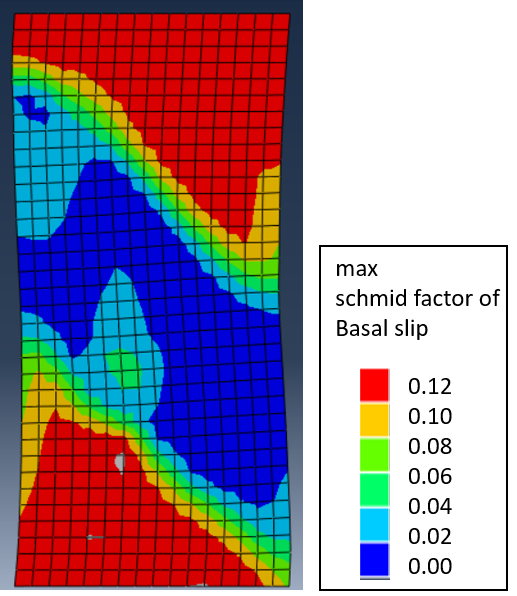}
\caption{Contour plot of the maximum local Schmid factor for basal slip in the twinned region and in the matrix for an applied compressive strain of 4\%.}
\label{fig:Schmid}
\end{figure}

The average shear strain per Gauss point accommodated by basal and pyramidal slip systems in the matrix and in the twinned region, according to the simulations, was plotted in Figure \ref{fig:slipMechanism} as a function of the applied compressive strain. As prismatic slips in both regions were very little compared other slip types, they are not included in the figure. In order to understand these results, it is interesting to know the highest Schmid factor for each slip system and tensile twin in the matrix and in the twinned region for a single crystal of Mg deformed in compression. They are depicted in Table \ref{tab:Schmid} and were calculated taking into account that the loading axis was parallel to the $[10\bar{1}0]$ direction in the matrix and changed to [01$\bar{1}$6] in the twinned region \citep{Wang2020}. After the initial elastic region, the simulation show that basal slip was activated in both the matrix and the twinned regions of the micro-pillar in the plateau region and it was more intense in the latter. The contour plot of the local values of maximum Schmid factor for basal slip in the matrix and in the twins is plotted in Figure \ref{fig:Schmid} for an applied strain of 4\%. The global Schmid factor in the twinned region (Table \ref{tab:Schmid}) is 0.13 but the local values are higher (reaching 0.2) in the selected zones. The global Schmid facor for basal slip in the matrix is 0 but slightly higher values are found locally (up to 0.05) near the twin/matrix interface. Thus, the activation of basal slip in the matrix can be explained because the CRSS for basal slip is very low and the local Schmid factor near the twin/matrix interface is different from zero. On the contrary, neither prismatic nor pyramidal were activated in the matrix and in the twin in the plateau region although the Schmid factors for both were high (0.49 and 0.39, respectively). Obviously, activation of these slip systems is hindered by the high values of their CRSS and the applied strain can be accommodated by the activation of twinning and basal slip at stresses below those necessary to activate either prismatic or pyramidal slip. 

\begin{table}[!tbp]
\centering%
\begin{tabular}{|l|l|l|}
\hline
 & Matrix & Twin \\ \hline
Basal &   0 & 0.13  \\ \hline
Prismatic & 0.46 & 0.01  \\ \hline
Pyramidal & 0.39 & 0.49   \\  \hline
Twin & 0.49 & --  \\
\hline
\end{tabular}%
\caption{Maximum values of the global Schmid factors for slip and twinning in the matrix and in the twinned regions of the micro-pillar loaded in compression.}
\label{tab:Schmid}
\end{table}

After the micro-pillar was fully twinned ($>$ 6\%), the applied strain was mainly absorbed by pyramidal slip, although there was also some contribution of basal slip when the compressive strains were below 7\%. The limited activity of basal slip in this regime (even though the Schmid factor was 0.13) can be explained because the basal plane in the twinned region is almost perpendicular to the loading axis and is not well oriented to accommodate the applied strain. However, basal slip bands were observed in the fully-twinned micro-pillar, in agreement with the predictions of the simulations. The Schmid factor for prismatic slip in the twinned region is very low and, thus, pyramidal slip is the main deformation mechanism, leading to a very large strain hardening. Again, these results are in good agreement with the experimental observations in \cite{Wang2020}. Thus, the CP-PF model was able to predict accurately the twin propagation and the activation of the different slip systems in the matrix and the twinned regions. It should be noted, however, that matrix/twin interface was diffusive and this representation is different from the sharp twin/matrix interface found experimentally.

\section{Simulation of twin growth in polycrystal}
\label{benchmarkPC}
\begin{figure}[!tbp]
\centering
\includegraphics[width=0.6\textwidth]{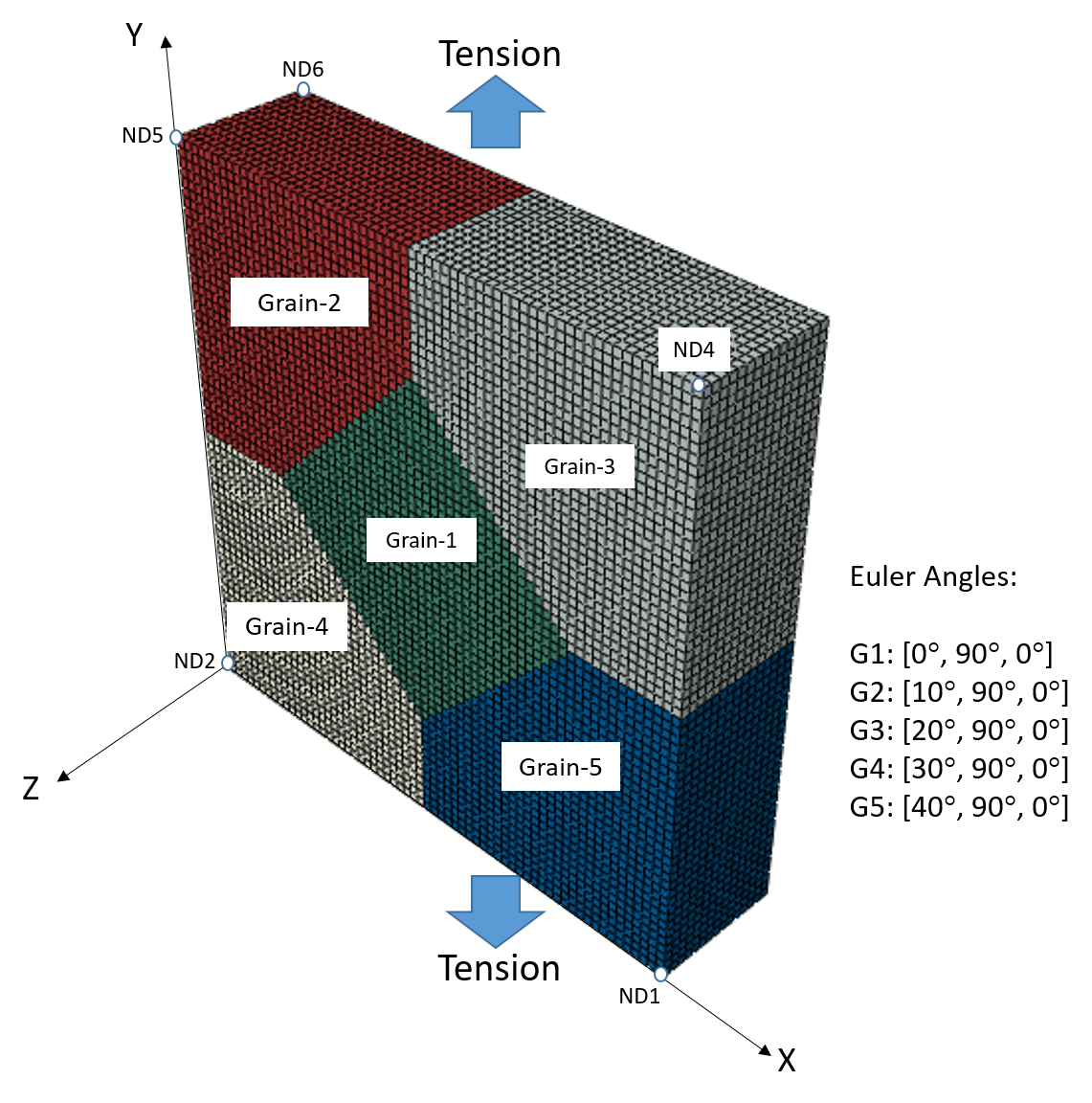}
\caption{Polycrystal containing a grain (grain-1) with favorable Schmid factor for twin growth. The Euler angles of each grain as well as the loading direction were indicated.}
\label{fig:meshPC}
\end{figure}

A column shaped polycrystal containing one grain favorably oriented for twin growth (tensile deformation is parallel to the c axis) was used to explore the capabilities of our CP-PF model. The initial orientations of the 5 grains are listed in Figure \ref{fig:meshPC} with Bunge type Euler angles.

The FEM domain contains $ 63 \times 63 \times 15 = 59535$ Abaqus C3D8R elements and the FFT domain has $ 64 \times 64 \times 16 = 65536$ cell points. Here the pad section of FFT grids was not used because the FEM domain is periodic along each direction. The four vertical surfaces were treated as free surfaces during the simulation. The vertical displacements of the bottom surface were impeded. Load was introduced by increasing the vertical displacement of all the nodes on the top surface of the polycrystal. The velocity of these nodes was set to achieve a strain rate of 10$^{-3}$ s$^{-1}$. Additionally, the degrees of freedom along $\mathbf{X}$ and $\mathbf{Z}$ in ND2 and ND5, along $\mathbf{X}$ in ND3 and ND6, as well as along $\mathbf{Z}$ in ND1 and ND4, were impeded.  

\begin{figure}[!tbp]
\centering
\includegraphics[width=0.8\textwidth]{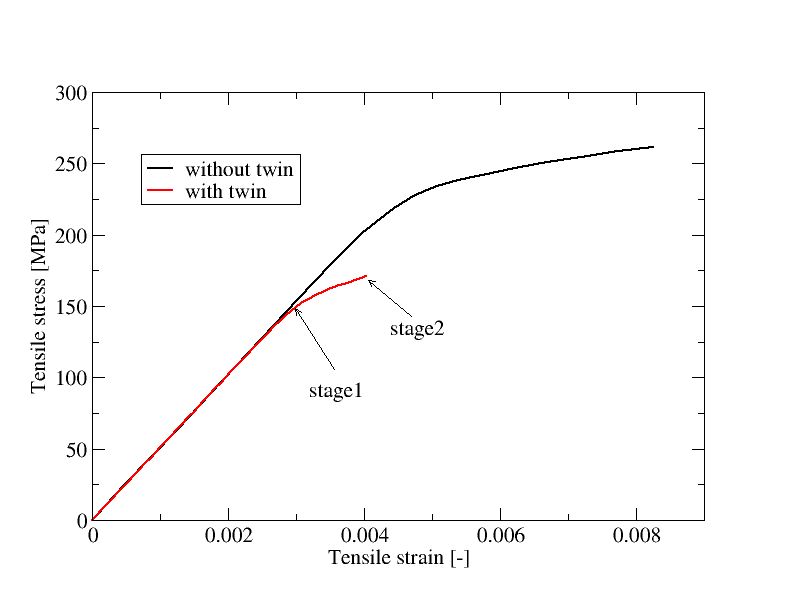}
\caption{Simulated stress-strain curves of polycrystal deformation with and without twinning. Two deformation stages to assess the twin geometry and the local stress are indicated with arrows in the stress-strain curves.}
\label{fig:sePlotPC}
\end{figure}

\begin{figure}[!tbp]
\centering
\includegraphics[width=1.0\textwidth]{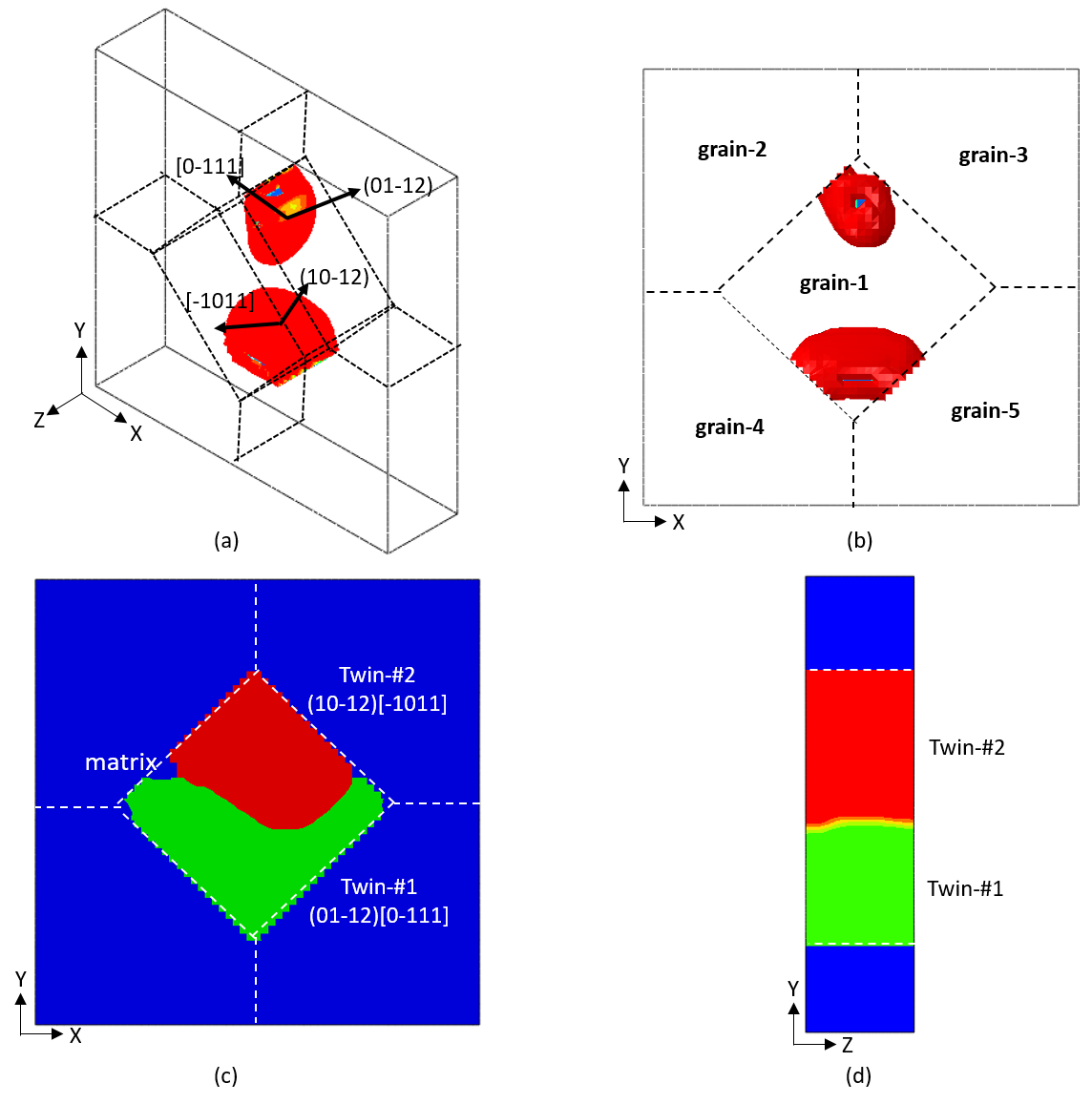}
\caption{Twin geometries at $\epsilon_{22}$=0.3\% (a,b) and $\epsilon_{22}$=0.4\% (c,d). Grain boundaries with dash lines as well as the normal direction and shear direction of twin variants are highlighted.  Two twin variants were marked in red in  (a) and (b), while in (c) and (d) the first and second twin variants were marked in green and red, respectively in (c) and (d). The 2D sections in (c) and (d) were obtained from the center of the simulation domain.}
\label{fig:twinGeo}
\end{figure}

The domain was strained along the $[0001]$ direction of Grain 1. The six twin variants have the same macroscopic Schmid factor (0.5) but only $(01\bar{1}2)[0\bar{1}11]$ and $(10\bar{1}2)[\bar0{1}11]$ twin variants were allowed to grow. The simulated global stress-strain curves with and without twinning are compared in Figure \ref{fig:sePlotPC}. Plastic deformation associated with the formation of twin occurs at 150 MPa in the polycrystal. As the macroscopic Schmid factor of twin is 0.5, the corresponding resolved shear stress $(10\bar{1}2)$ plane along $[\bar{1}011]$ direction is 75 MPa. This value is not far away from measured CRSS of twin in Figure \ref{fig:expSE}. The geometry of the twin variants at $\epsilon_{22}$=0.3\% and $\epsilon_{22}$=0.4\% (\emph{stage1} and \emph{stage2} in Figure \ref{fig:sePlotPC}) are plotted in Figure \ref{fig:twinGeo}. It was found that the main growth directions of the two twin nuclei were different. Both twins have plate shape and the thickness of twin along twin normal direction was clearly smaller than the twin length along the $(10\bar{1}2)$ plane. Twins have engulfed the whole grain1 at $\epsilon_{22}$=0.44\% and the twin-twin interface is not flat is made up of sections parallel to the nearest grain boundary. 

\section{Discussion}

\label{DisSizeMap}

The uniqueness of the current numerical strategy comes from combination of a low-resolution finite element mesh for the CP simulation with a high-resolution regular grid for phase field evolution using FFT. The numerical efficiency and accuracy of this strategy depends on the density of FFT cells and on the FEM-FFT mapping approach and both topics are discussed below.

\subsection{Influence of discretization}

The influence of the discretization on the accuracy of the twin evolution was assess using the micro-pillar compression test as an example. Three different discretizations \emph{mesh1}, \emph{mesh2} and \emph{mesh3} were used:

\begin{itemize}
	\item \emph{mesh1:} The number of integration points in the FEM (8704) is comparable to the total number of FFT grids (16x16x32=8192).  
	\item \emph{mesh2:} The number of FFT grids (32x32x64=65536) is almost 8 times higher than the number of integration points in the FEM model (8704).
	\item \emph{mesh3:} The number of FFT grids (64x64x128=524288) is almost 64 times higher that the number of integration points in the FEM model (8704).
\end{itemize}

The twin boundaries in FFT domain in the reference configuration were plotted when the applied compressive strain of the micropillar was 0.4 for the three different discretizations. The coarser meshes (\emph{mesh1} and \emph{mesh2}) are not able to predict that the coherent twin interface parallel to $(10\bar{1}2)$ plane, which is the experimental result in Figure \ref{fig:expDetail}(a). On the contrary, the finer mesh is able to predict the actual twin interface orientation. The mesh density also has a large influence on the flow stress, which was much lower in \emph{mesh3} as compared with \emph{mesh2} for the same set of material parameters.

\begin{figure}[!tbp]
\centering
\includegraphics[width=1.0\textwidth]{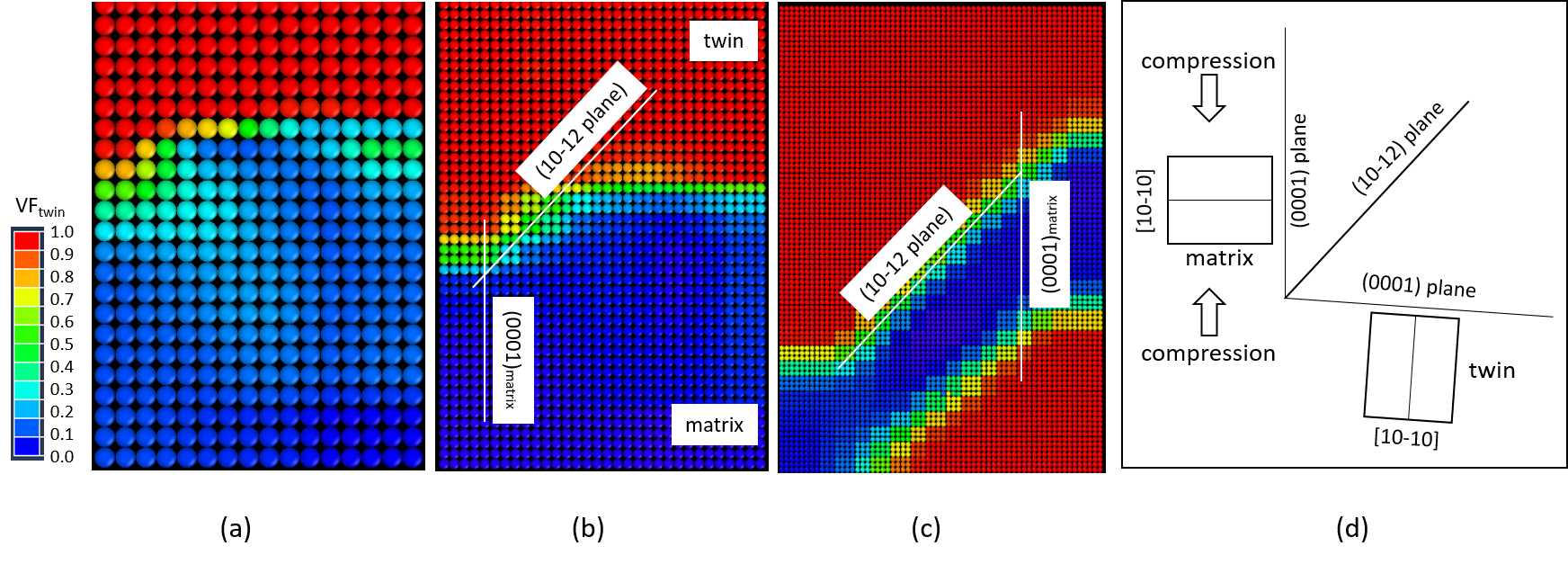}
\caption{Contour plot of the phase field variable showing the twin-matrix interfaces predicted with same FEM discretization but different FFT grids when the compressive applied strain in the micro-pillar was 0.4. Only a small part of boundaries are parallel to the coherent twin interface in \emph{mesh1} (a) and \emph{mesh2} (b), while most of the interfaces are parallel to $(10\bar{1}2)$ in \emph{mesh3} (c). (d) Schematic  of the orientation of the coherent twin interface. The FEM mesh is same as Figure \ref{fig:elements}(a).}
\label{fig:compareMeshSize}
\end{figure}

\subsection{Comparison of different mapping approaches}

The simulated matrix-twin interface is also sensitive to the coupling way between FEM mesh and FFT cell. The advantages and disadvantages of three different mapping strategies are compared below. 

\begin{itemize}
	\item \emph{benchmark:} The simulation results of section \ref{benchmarkPC} serves as benchmark because deformation and phase field evolution problems were solved using the same regular cubic mesh and  mesh mapping is not an issue.  
	\item \emph{VoronoiMap:} In this fast FEM-FFT mapping, the FFT cell and FEM Gauss point with the shortest distance between them form pairs, and the data between them were transferred directly. 
	\item \emph{interpolation} A subdomain of the Gauss points in the FEM is constructed for each FFT cell. The stresses at Gauss points are evaluated with specific weighting factors and assigned to the FFT cell. The subdomain of FFT cells  is constructed for each Gauss point, and the twin volume fractions of FFT cells are evaluated with weighting factors and assigned to the Gauss point.
\end{itemize}

For the \emph{interpolationMap}, a subdomain for each Gauss point inside FFT cell as well as a subdomain for each FFT cell inside FEM Gauss points is initially selected. Then, the source rate is transferred from FEM mesh to FFT cell and the phase field value from FFT cell is transferred to FEM mesh as follows
\begin{equation}
r_{\alpha}(\mathbf{x}^{\prime}) = \sum^{N}_{I} r_{\alpha}(\mathbf{x}_{I})W(\mathbf{x}_{I}-\mathbf{x}^{\prime}) 
\label{avgMap1}
\end{equation}
\begin{equation}
\boldsymbol{\phi}(\mathbf{x}) = \sum^{N}_{I} \boldsymbol{\phi}(\mathbf{x}^{\prime}_{I})W(\mathbf{x}^{\prime}_{I}-\mathbf{x}) 
\label{avgMap2}
\end{equation}

\noindent with the weighting factors

\begin{equation}
W(\mathbf{x}_{I}-\mathbf{x}^{\prime})=\frac{1}{\left\| \mathbf{x}_{I}-\mathbf{x}^{\prime} \right\|} \left( \sum^{N}_{J} \frac{1}{\left\| \mathbf{x}_{J}-\mathbf{x}^{\prime} \right\|} \right)^{-1}
\label{avgMap3}
\end{equation}

\noindent
where $\mathbf{x}^{\prime}$ and $\mathbf{x}$ are the location of FFT cell and FEM Gauss points, respectively. The interpolation was carried out within Gauss points or FFT cells in this example.

\begin{figure}[!tbp]
\centering
\includegraphics[width=1.0\textwidth]{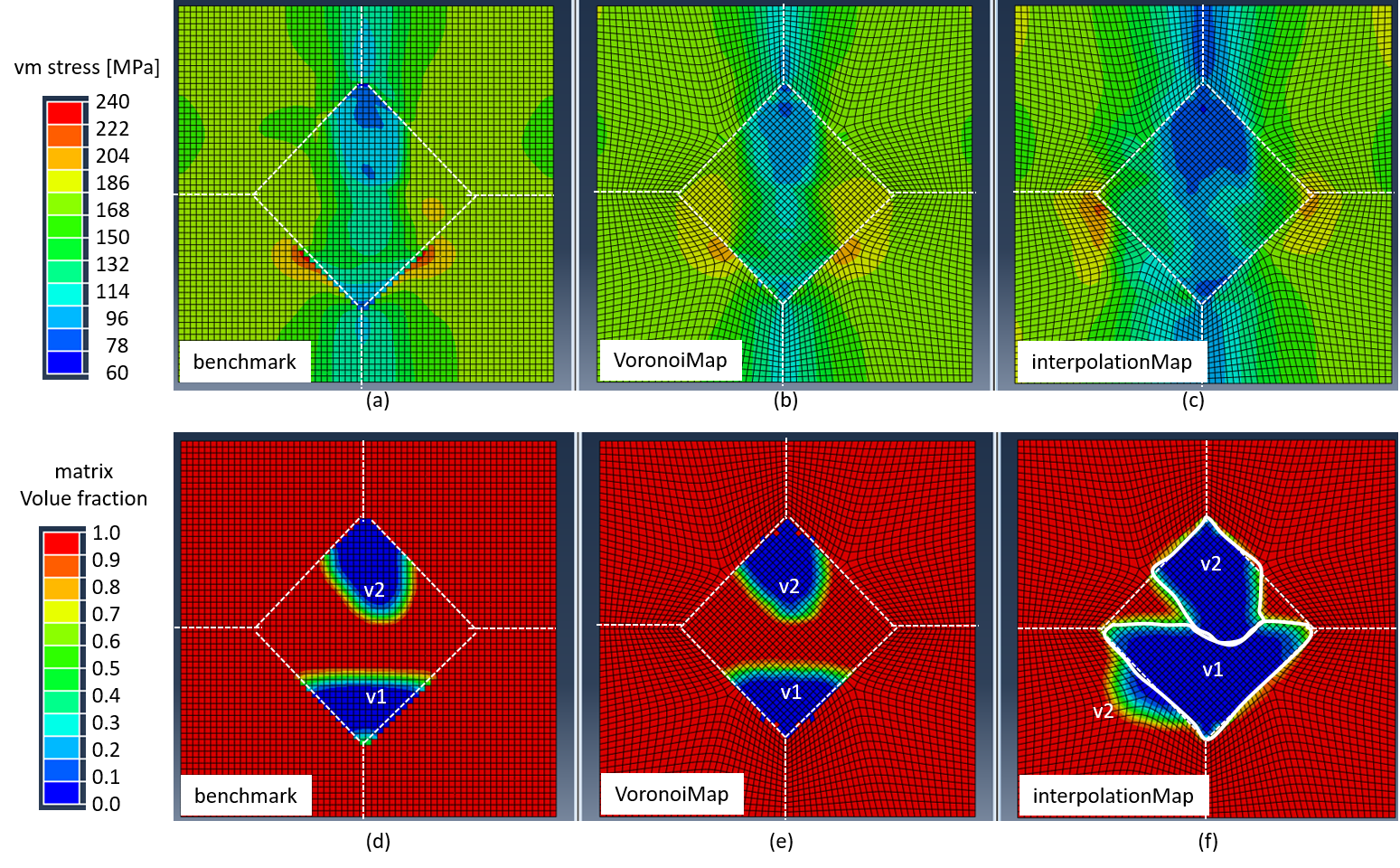}
\caption{Comparison of von Mises stresses and of the matrix volume fraction predicted with different FEM-FFT mapping strategies. The \emph{VoronoiMap} method predicted similar results as \emph{benchmark} while the \emph{interpolationMap} method overshooted the interface motion and showed unexpected penetration through the grain boundary. The phase boundaries between variants V1 and V2 as well as the penetration caused V2 at the left bottom grain of (f) are highlighted.}
\label{fig:compareAvg}
\end{figure}

The contour plots of the von Mises stress and of the matrix volume fraction are plotted in Figures in Figure \ref{fig:compareAvg} (a) to (c) and \ref{fig:compareAvg} (d) to (f), respectively, for the three different interpolation strategies (\emph{benchmark}, \emph{VoronoiMap} and \emph{interpolationMap}). The results of the simulations show that the Voronoi mapping strategy used in our model provides very close results to the benchmark from the viewpoint of the stress fields and twin evolution. Surprisingly, the widely used \emph{interpolationMap} brought unexpected results, i.e. overshooting interface motion and phase field flux running across grain boundary.

Thus, modelling with a low-resolution FEM mesh to account for plastic slip and a high-resolution regular grid for phase field evolution to account for twining using the \emph{VoronoiMap} is a valid strategy to keep a balance between computational efficiency and accuracy of the results.

\section{Conclusions}
A numerical strategy to simulate plastic deformation in Mg alloys including dislocation slip and twin propagation has been presented. Dislocation slip is included through a crystal plasticity model that was solved using the finite element method. Twin propagation is taken into account by means of a phase field model that was solved using a FFT algorithm. The CP model and PF model were solved using different discretizations of the simulation domain but the same time step. At each time step the numerical strategy always began with solving of the CP model. Then calculated variables at FEM integration points were transferred to the FFT cells where PF model governing equations were manipulated. Finally the updated twin volume fractions at each FFT cell were transferred back to the FEM integration points. The uniqueness of the current numerical strategy is that a low-resolution finite element mesh for deformation and a high-resolution regular grid for phase field evolution with an efficient FEM-FFT mapping strategy were proposed and implemented.  

The numerical strategy was used to simulate the compression of a Mg micro-pillar along the $[10\bar{1}0]$ direction, as well as plastic deformation and twin growth inside a Mg polycrstal. The model predicted the propagation of a tensile twin along the micro-pillar at constant stress followed by a strong strain hardening after the micro-pillar was fully twinned, in excellent agreement with the experimental results in the literature \citep{Wang2020}. In addition, the numerical model predicted that basal slip was active in the matrix near interface and in the twin region during twin propagation, while the pyramidal slip was the dominant deformation mechanism in the fully-twinned micro-pillar, again in agreement with the experimental results. Additionally, multi-variants twining inside a polycrystal were simulated to show the capabilities of the model.

\section{Acknowledgements}

This investigation was supported by the European Union Horizon 2020 research and innovation programme (Marie Sklodowska-Curie Individual Fellowships, Grant Agreement 795658) and the Comunidad de Madrid Talento-Mod1 programme (Grant Agreement PR-00096). Additional support from the European Research Council under the European Union's Horizon 2020 research and innovation programme (Advanced Grant VIRMETAL, grant agreement No. 669141) and by the HexaGB project of the Spanish Ministry of Science and innovation (with reference number RTI2018-098245, MCIN/AEI/10.13039/501100011033/) is gratefully acknowledged.


\end{document}